\documentclass[amscd,amssymb,verbatim]{amsart}
\usepackage{epsfig}
\usepackage{graphics}
\usepackage{newlfont}
\theoremstyle{plain}
\newtheorem{Thm}{Theorem}
\begin{document}
\numberwithin{equation}{section}
 \title{Taming of the Wild Group of Magnetic Translations} 
\author{Peter Varga}
\address{Institute of Mathematics and Informatics \\
  Lajos Kossuth University\\
  H-4010 Debrecen, Hungary} \email{varga@@math.klte.hu} \thanks{}
\keywords{Solvable groups, Type II representations, quasi-periodic
 systems, solid-state physics } 
\date{November 27, 1996}

\begin{abstract}
We use a theorem of Auslander and Kostant on the representation
theory of solvable Lie-groups for the study of some groups necessary
for the description of certain quasi-periodic systems of solid-state
physics. We show that the magnetic translation group is tame (Type I) if the
magnetic field is not constant but fluctuating.
\end{abstract}

\maketitle

\section{Introduction} \label{S:intro}

Lie-groups are divided into two classes (Types I and II) according to
the behaviour of their representations \cite{Mackey}. The unitary
representations of Type I (tame) groups have essentially  unique
 decompositions into
 irreducible representations, while in the case of
Type II (wild) groups such decomposition can be highly
nonuniqe. Finite groups, semisimple and nilpotent Lie-groups are tame,
while infinite discrete groups (except those which contain  an Abelian
subgroup of finite index) are wild. The type of a solvable Lie-group
  is determined by the behaviour of its coadjoint orbits. According to a
theorem of Auslander and Kostant \cite{AK}, a solvable Lie-group
 is tame if and
only if the set of its coadjoint orbits are separable and the their standard
symplectic two-forms are exact. This theorem provides a fairly
convenient method to prove the wildness of some solvable groups.
The notation of Type I and II representations comes from the theory of
 von Neumann algebras. This operator algebraic aspect might be
 especially relevant in physical applications, where one is interested
 in the properties of the representations of the enveloping algebra. 
However, we have little to say about this topics in the present paper.

In Kirillov's book \cite{Kirillov} two simple examples of wild
solvable groups are given. These examples are not just mathematical
curiosities, but they emerge naturally in the description of some
quasi-periodic systems in solid-state physics.  Kirillov's first
example has the following physical realization: The functions $
a\cos{x}$, $a\sin{x}$, $b\cos{\alpha x}$, 
$b\sin{\alpha x}$, and the derivation
$\partial_{x}$ form a five dimensional Lie-algebra. If $\alpha$ is
irrational, then  its Lie-group is wild. These operators are the
building blocks of the Hamiltonian of an electron moving  a 
quasi-periodic  cosine potential.

The Lie-algebra of the second example can be represented by operators
which are necessary for the description of  the motion of an electron
in two dimension under the influence of periodic cosine potentials and
uniform magnetic field. The corresponding group contains the magnetic
translation group \cite{Zak,Hofs}.

The physics of quasi-periodic systems has many characteristic features
like the unusual band structure, various types of (de)localisations,
etc. \cite{Sokoloff}. The wildness of the groups in these examples
foreshadows the appearance of such features, so the theorem of
Auslander and Kostant can be used to predict the qualitative nature of
physical systems connected with solvable Lie-groups. In particular, we
show that if the Lie-algebra of the magnetic translation group is extended by
generators generating fluctuations of the magnetic field, then the
corresponding Lie-group becomes tame, so in that case the unusual
fractal band structure is not expected.

In the next section we re-present the examples of \cite{Kirillov} and
give physical realizations of the wild solvable groups. We also
determine how the characters of the systems changes if some parameters
like the magnitude of the potential and magnetic field is allowed to
fluctuate. This paper is basically an extra exercise for the last
section of \cite{Kirillov}.

\section{Solvable Lie-groups in solid-state
physics.}\label{S:solvable}

Let us first recall the notation of coadjoint orbits. Let $G$ be a
Lie-group, $\mathfrak{g}$ its Lie-algebra, and $\mathfrak{g}^\ast$ its dual.
The coadjoint action of $G$ on $\mathfrak{g}^\ast$ is defined by
\begin{equation}\label{E:co1}
\langle Ad^{\ast}_{g}\xi, Ad_{g} X \rangle
=\langle{}\xi{},X\rangle{},\quad 
\xi \in \mathfrak{g}^{\ast}, X \in \mathfrak{g},g \in{}G.
\end{equation}
By differentiating \eqref{E:co1} we obtain
\begin{equation}\label{E:co2}
\langle ad^{\ast}_{X}\xi,Y\rangle = -\langle\xi,[X,Y]\rangle.
\end{equation}
On the orbits ${\Omega}_{\xi_{0}}=
\left\{
  Ad^{\ast}_{g}\xi_{0},g\in G
\right\}
$
$\,ad^{\ast}_{X}$ is represented by a vector field
$f_{\Omega_{\xi_{0}}}(X)$ . The symplectic two-form $B_{\Omega}$ on
$\Omega$ is given by 
\begin{equation}\label{E:co3}
B_{\Omega_{\xi}}
\left(f_{\Omega_{\xi}}(X),f_{\Omega_{\xi}}(Y)\right)(\xi)=
\langle\xi,[X,Y]\rangle.
\end{equation}
A theorem of Auslander and Kostant characterizes the simply connected
solvable Type I Lie-groups:
\begin{Thm}\label{T:AK}
Let $G$ be a simply connected solvable Lie-group. Then G is Type I
(tame) if and only if
\begin{enumerate}
\item  
 all coadjoint orbits of $G$ are $G_{\delta}$ sets (i.e. they
  are countable intersections of open sets) in the usual topology on
  $\mathfrak{g}$ .
\item 
 The symplectic forms $B_{\Omega_{\xi}}$ are exact for all 
$\xi\in \mathfrak{g}^{\ast}$.
\end{enumerate}
\end{Thm}
We use this theorem for the study of some Lie-groups connected with
the theory of quasi-periodic systems in solid-state physics.

The simplest example of wild groups is the five dimensional Mautner
group \cite{Kirillov} consisting of certain $3\times 3$ complex
matrices:
\begin{equation}\label{E:e1}
g(t,w,z)=
\begin{pmatrix}
e^{it} & 0             & z \\
0      & e^{i\alpha t} & w \\
0      & 0             & 1
\end{pmatrix},
\quad t \in \Bbb{R},\,  z,w \in \Bbb{C},
\end{equation}
where $\alpha$ is a fixed irrational number.
The non-zero commutators of the Lie-algebra of this group are
\begin{equation} \label{E:e2}
\begin{aligned}
\mbox{}& [P,S_{1}]=C_{1}, \\
& [P,C_{1}]=-S_{1},
\end{aligned}
\text{\mbox{}}\qquad\qquad
\begin{aligned}
\mbox{}&[P,S_{\alpha}]=\alpha C_{\alpha}, \\
&[P,C_{\alpha}]=-\alpha S_{\alpha}.
\end{aligned}
\end{equation}

Operators satisfying the same algebra occur in the theory of
one-dimensional quasi-periodic systems. A representation of
\eqref{E:e2} is provided by the following operators acting on
$L^{2}(\Bbb{R},dt)$:
\begin{equation}\label{E:e3}
\begin{aligned}
P=\partial_{t},
\end{aligned}
\text{\mbox{}}\qquad
\begin{aligned}
\mbox{}& S_{1}=a\sin{(t+\phi_{1})},\\
&C_{1}=a\cos{(t+\phi_{1})},
\end{aligned}
\text{\mbox{}}\qquad
\begin{aligned}
\mbox{}& S_{\alpha}=a_{\alpha}\sin{(\alpha t+\phi_{\alpha{}})},\\
&C_{\alpha}=a_{\alpha}\cos{({\alpha}t+\phi_{\alpha{}})}.
\end{aligned}
\end{equation}
A representation with different
$a_{1}',a_{\alpha{}}',\phi{}_{1}',\phi{}_{\alpha{}}'$ parameters is
isomorphic to \eqref{E:e3} iff
$a_{1}=a_{1}',a_{\alpha{}}=a_{\alpha{}}' $ and 
$\phi{}_{1}-\phi{}/{\alpha{}}=
\phi{}_{1}'-\phi{}'/{\alpha{}}+2m\pi{}+2n\pi{}/\alpha{}$ for some 
$m,n \in{} \Bbb{Z}$.
One can build the
Hamiltonian of an electron moving in a quasi-periodic cosine potential
out of these operators:
\begin{equation}\label{E:e4}
  H=-\frac{1}{2}\partial_{x}^{2}{}
     +a_{1}\cos{(t+\phi_{1})}+a_{\alpha}\cos{(\alpha
  t+\phi_{\alpha})}=
  -\frac{1}{2}P^{2}+a_{1}C_{1}+a_{\alpha}C_{\alpha}.
\end{equation}

In \cite{Kirillov} two inequivalent decompositions of the regular
representation of \eqref{E:e1} into irreducible ones  are presented.
Inequivalent decompositions of a representation of \eqref{E:e2}
occurred in the physics literature, too.
It was noted in \cite{Romerio,Wollf,WJJ,JJ} that although \eqref{E:e4} has no
translational symmetry, it is not completely random either. By adding
an extra dimension, translations by $2\pi$ and $2\pi/\alpha$ can be
executed in separate dimensions. For that purpose, we consider the
following representation of \eqref{E:e2} on $L^{2}(\Bbb{R}^{2},dxdy)$:
\begin{equation}\label{E:e5}
\begin{aligned}
P=\partial_{x}+\partial_{y}
\end{aligned}
\text{\mbox{}}\qquad
\begin{aligned}
\mbox{}& S_{1}=a_{1}\sin{x},\\
&C_{1}=a_{1}\cos{x},
\end{aligned}
\text{\mbox{}}\qquad
\begin{aligned}
\mbox{}& S_{\alpha}=a_{\alpha}\sin{\alpha y},\\
&C_{\alpha}=a_{\alpha}\cos{\alpha y}.
\end{aligned}
\end{equation}
Since $P$ is the generator of translations only along the lines 
$l_{c}:\,y=x+c$, the representation \eqref{E:e7} is decomposable into
irreducible representations acting on the Hilbert-spaces
$L^{2}(l_{c})$. These representations are isomorphic to \eqref{E:e3}
with parameters $\phi_{1}{}=0,\,\phi{}_{\alpha{}}=c\alpha{}$.
A different decomposition of $L^{2}(\Bbb{R}^{2},dxdy)$ is based on the
periodicity of \eqref{E:e5} on the $xy$-plane.
The operator $H=-1/2P^{2}+a_{1}C_{1}+a_{\alpha}C_{\alpha}$ is indeed invariant
against the translations $(x,y)\to (x+2\pi,y)$ and 
$(x,y) \to (x,y+2\pi/\alpha)$. The translational symmetry entails the
existence of Bloch wave-functions 
\begin{equation}\label{E:e6}
\psi(x+2\pi,y)=e^{is}\psi(x,y) ,\qquad\qquad
\psi(x,y+\frac{2\pi}{\alpha})=e^{it}\psi(x,y).
\end{equation}
The operators acting on such wave-functions for fixed $s$ and $t$
provide exactly the infinitesimal form of the irreducible
representation occurring in the second decomposition of the regular
representation in \cite{Kirillov}. Indeed, if we introduce the
periodic functions
\begin{equation}\label{E:e7}
\tilde{\psi}(x,y)=e^{-i(sx+t\alpha y)}\psi(x,y),
\end{equation}
then the operators \eqref{E:e5} act on $\tilde{\psi}$ as
\begin{equation}\label{E:e8}
\tilde{P}=\partial_{x}+\partial_{y}+i(s+\alpha t), \qquad\,
(S_{1},C_{1},S_{\alpha},C_{\alpha} \,\,\mbox{are unchanged}).
\end{equation}
Since $\tilde{\psi}$ is periodic, we can regard it as a function
defined on the torus \\
 $S^{1}\times S^{1}=[0,2\pi)\times[0,2\pi/\alpha)$.
The action of the operators \eqref{E:e8}  on $L^{2}(S^{1}\times{}S^{1})$  is
irreducible. 
The existence of a representations with Bloch wave-functionals does
not {\em{a priori}} implies the occurance  of  extended states in the
physical representation \eqref{E:e5}. Indeed, as it was stressed in
\cite{Bellissard}, inequivalent representations of the same algebra
might have very different spectral and localizational properties.
Nevertheless, the existence of extended states in this system was
established in \cite{BLT,Si,FSW}.

Next we study the effect of the fluctuation of the magnitude of the
potential. For this purpose we add the generator $M=\partial_{a}$ to
the operators of \eqref{E:e3}. $M$ changes the amplitudes of the
potentials $S_{1}$ and $C_{1}$. To keep the algebra closed we need to add the
operators $S_{0}=sin{(t+\phi_{1})}$ and $C_{0}=cos{(t+\phi_{1})}$ to
\eqref{E:e3}, too. The extra non-zero commutators (compared to
\eqref{E:e2}) of the extended Lie-algebra $\mathfrak{g}$ are
\begin{equation}\label{E:e8b}
[P,S_{0}]=C_{0},\quad [P,C_{0}]=-S_{0},\quad
[M,S_{1}]=S_{0},\quad [M,C_{1}]=C_{0}.
\end{equation}
The Lie-group $G$ of $\mathfrak{g}$ has a representation by $4\times
4$ matrices
\begin{equation}\label{E:e9}
g(t,a,u,w,z)=
\begin{pmatrix}
e^{it} &  a      &   0              &   u  \\
0      &  e^{it} &   0              &   z  \\
0      &  0      &   e^{i\alpha t}  &   w  \\
0      &  0      &   0              &   1
\end{pmatrix},
\qquad a,t\in \Bbb{R}, \,\, u,w,z\in \Bbb{C}.
\end{equation}
If $\mathfrak{g}^{\ast}$ is represented by matrices of the following
form
\begin{equation}\label{E:e10}
\xi(\tau,p,l,m,)=
\begin{pmatrix}
i\tau & 0 & 0 & 0 \\
p     & 0 & 0 & 0 \\
0     & 0 & 0 & 0 \\
l     & m & n & 0
\end{pmatrix},
\qquad \tau,t \in \Bbb{R},\,\, l,m,n \in \Bbb{C},
\end{equation}
so the pairing between $\mathfrak{g}$ and $\mathfrak{g^{\ast{}}}$ is
\begin{equation}\label{E:10a}
\langle \xi{},h \rangle{}=
\Re{\left({}\operatorname{Tr}(\xi{}h)\right){}},\quad{}
\xi{}\in{}\mathfrak{g^{\ast{}}},h\in{}\mathfrak{g},
\end{equation}
then the coadjoint action is
\begin{multline}\label{E:e11}
Ad^{\ast}_{g(t,a,u,w,z)}\xi
\left(\tau,p,l,m,n\right)=\\
\xi\left(\tau+\Im{(lu+zm+\alpha nw)}, p-\Re{(lz)},
le^{-it},me^{-it}-\Re{(lz)},ne^{-i\alpha t}\right).
\end{multline}
The four dimensional orbits are given by the parametric equations
\begin{equation}\label{E:e12}
l=l_{0}e^{it}, \qquad n=n_{0}e^{i\alpha t}.
\end{equation}
Since the orbits are dense subsets of the sets
\begin{equation}\label{E:e13}
|l|=l_{0},\qquad |n|=n_{0}
\end{equation}
the first criteria of the Auslander-Kostant theorem fails, so the
group remains wild despite the fluctuation of the potential. 

In the following we turn our attention to Kirillov's second example of
wild groups. 
This group is closely related to the magnetic translation group, whose
Type II nature at irrational magnetic flux was pointed out by \cite{Gr}
This is a seven dimensional Lie-algebra whose nonzero
commutators are
\begin{equation}\label{E:f1}
\begin{aligned}
\mbox{}[P_{x},P_{y}]=2B,
\end{aligned}
\text{\mbox{}}\qquad
\begin{aligned}
\mbox{}& [P_{x},S_{x}]=C_{x},\\
&[P_{x},C_{x}]=-S_{x},
\end{aligned}
\text{\mbox{}}\qquad
\begin{aligned}
\mbox{}& [P_{y},S_{y}]=C_{y},\\
&[P_{y},C_{y}]=-S_{y}.
\end{aligned}
\end{equation}
This algebra is represented by the operators
\begin{equation}\label{E:f1a}
\begin{aligned}
\mbox{}& \hat{P_{x}}=i\partial_{x}-by, \\
 & \hat{P_{y}}=i\partial_{y}+bx,
\end{aligned}
\text{\mbox{}}\qquad
\begin{aligned}
\mbox{}& \hat{C_{x}}=\cos{x},\\
& \hat{S_{x}}=\sin{x},
\end{aligned}
\text{\mbox{}}\qquad
\begin{aligned}
\mbox{}&\hat{C_{y}}=\cos{y} ,\\
& \hat{S_{y}}=\sin{y},
\end{aligned}
\text{\mbox{}}\qquad
\begin{aligned}
\hat{B}=b,
\end{aligned}
\end{equation}
on $L^{2}(\Bbb{R}^{3},dx\,dy\,dz)$. The Hamiltonian of an electron
moving in constant magnetic field in a periodic crystal can be formed
out of these operators:
\begin{equation}
\hat{H}=\hat{P_{x}}^{2}+\hat{P_{y}}^{2}+\hat{C_{x}}+\hat{C_{y}}.
\end{equation}

If we regard the generators as linear functions on
$\mathfrak{g}^{\ast}$, then the coadjoint orbits are
\begin{equation}\label{E:f2}
C_{x}^{2}+S_{x}^{2}=r_{1}^{2}, \qquad
C_{y}^{2}+S_{y}^{2}=r_{2}^{2}, \qquad  
B=r_{3}.
\end{equation}
If the orbits are parametrized as 
\begin{equation}\label{E:f3}
C_{x}=r_{1}\cos{\phi},\qquad S_{x}=r_{1}\sin{\phi},\qquad
C_{y}=r_{2}\cos{\psi},\qquad S_{y}=r_{2}\sin{\psi},
\end{equation}
then the symplectic two-form $B_{\Omega}$ is
\begin{equation}\label{E:f4}
B_{\Omega}=d\phi\wedge dP_{x} +d\psi\wedge dP_{y} 
+2r_{3}d\phi \wedge d \psi.
\end{equation}
Since
\begin{equation}\label{E:f5}
\int \limits_{\{P_{x}=P_{y}=0,\,B=r_{3}\}} B_{\Omega} = 8\pi^{2}r_{3},
\end{equation}
$B_{\Omega}$ is not exact, so the second criteria of the
Auslander-Kostant theorem fails, consequently the group of magnetic
translations is wild.

Now let us see what happens if the external magnetic field is
dynamical, too. To describe the fluctuation of $b$ we extend the set
of generators \eqref{E:f2} by $\hat{E}=i\partial_{b}$. In order to
keep the commutators closed, we need to adjoin the operators
$\hat{Y}=-i[\hat{E},\hat{P_{x}}]$ and $\hat{X}=i[\hat{E},\hat{P_{y}}]$,
too. So the following eleven-dimensional Lie-algebra is necessary to
describe the coupled system of an electron and the fluctuating
external magnetic field:
\begin{equation}\label{E:f6}
\begin{aligned}
\mbox{}&  [P_{x},S_{x}]=C_{x}    , \\
 &         [P_{x},C_{x}]=-S_{x}    ,
\end{aligned}
\text{\mbox{}}\qquad
\begin{aligned}
\mbox{}&  [P_{y},S_{y}]=C_{y}          ,\\
&        [P_{y},C_{y}]=-S_{}           ,
\end{aligned}
\text{\mbox{}}\qquad
\begin{aligned}
\mbox{}&  [P_{x},X]=I    ,\\
&        [P_{y},Y]=I   ,
\end{aligned}
\end{equation}
\begin{equation}\label{E:f66} \notag
\begin{aligned}
\mbox{}&  [E,P_{x}]=-Y   , \\
 &         [E,P_{y}]=X   ,
\end{aligned}
\text{\mbox{}}\qquad
\begin{aligned}
\mbox{}&  [P_{x},P_{y}]=2B          ,\\
&        [E,B]=I           . \notag
\end{aligned}
\end{equation}
If we use the generators of the Lie-algebra as linear functions on
$\mathfrak{g}^{\ast}$ then the coadjoint action  corresponds to the
following vector fields:
\begin{align}\label{E:f7}
\mbox{}  & V_{P_{x}}=-C_{x}\partial_{S_{x}}+S_{x}\partial_{C_{x}}+
  2B\partial_{P_{y}}+Y\partial_{E}+I\partial_{Y}, \notag \\
  & V_{P_{y}}=-C_{y}\partial_{S_{y}}+S_{y}\partial_{C_{y}}-
  2B\partial_{P_{x}}-X\partial_{E}+I\partial_{X}, 
\end{align} 
\begin{equation}\label{E:f8}
  \begin{aligned}
\mbox{}    & \phantom{V_{S_{x}}} V_{S_{x}}=C_{x}\partial_{P_{x}}, \\
    & \phantom{V_{S_{x}}} V_{C_{x}}=-S_{x}\partial_{P_{x}}, \\ 
    & \phantom{V_{S_{x}}} V_{B}=-I\partial_{E},  \\
    & \phantom{V_{S_{x}}} V_{X}=-I\partial_{P_{x}},
  \end{aligned}
  \text{\qquad \phantom{and} \qquad}
  \begin{aligned}
\mbox{}    & V_{S_{y}}=C_{y}\partial_{P_{y}}, \\
    & V_{C_{y}}=-S_{y}\partial_{P_{y}}, \\
    & V_{E}=I\partial_{B}-Y\partial_{P_{x}}+X\partial_{P_{y}}, \\
    & V_{Y}=-I\partial_{P_{y}},
  \end{aligned} 
\end{equation}
\begin{equation}\label{E:f9}
  V_{I}=0. 
\end{equation}
Note that $\partial_{I}$ does not occur in these expressions, so
$I=I_{0}=const.$ on each orbit. The form of $V_{P_{x}}$ and
$V_{P_{y}}$  implies that
\begin{equation}\label{E:f10}
C_{x}^{2}+S_{x}^{2}=r_{x}^{2}, \qquad \qquad 
C_{y}^{2}+S_{y}^{2}=r_{y}^{2}, 
\end{equation}
while \eqref{E:f8} entails 
\begin{equation}\label{E:f10b}
\cal{L}\left(\{V_{X},V_{Y},V_{E},V_{B}\}\right)=
 \cal{L}\left(\{\partial_{P_{x}},\partial_{P_{y}},
                \partial_{E},\partial_{B}\}\right).
\end{equation}
 So the orbits are
generated by the vectors $\partial_{P_{x}},\partial_{P_{y}},
\partial_{E},\partial_{B}$ and by
\begin{equation}\label{E:f11}
\tilde{V}_{P_{x}}=-C_{x}\partial_{S_{x}}
                  +S_{x}\partial_{C_{x}}+I_{0}\partial_{Y},
\qquad
\tilde{V}_{P_{y}}=-C_{y}\partial_{S_{y}}
                  +S_{y}\partial_{C_{y}}+I_{0}\partial_{x}.
\end{equation}
The integral manifolds of these vectors are
\begin{align}\label{E:f12}
 \mbox{} & C_{x}=r_{x}\cos{\phi}, \qquad
           S_{x}=r_{x}\sin{\phi}, \qquad
           Y=I_{0}(\phi+\phi_{0}) \notag \\
 \mbox{} & C_{y}=r_{y}\cos{\psi}, \qquad
           S_{y}=r_{y}\sin{\psi}, \qquad
           X=I_{0}(\psi+\psi_{0}),
\end{align}
while $E,B,P_{x},P_{y}$ are arbitrary.  So the maximal dimensional
orbits are homeomorph to $\Bbb{R}^{6}$. Since $H^{2}(\Bbb{R}^{6})=0$,
the symplectic two-form $B_{\Omega}$ is necessarily
exact.  Consequently this extension of the magnetic translation group
is tame!
\vspace{24pt}

{\bf Acknowledgements:}
We are grateful to Prof. Istv\'{a}n Kov\'{a}cs for  discussions
on operator algebras. We thank  the  referee for helpful suggestions and
for calling our attention to some references. This research was
partially funded by the grant OTKA-F015470.

\end{document}